\documentclass[preprint,showpacs,showkeys,superscriptaddress,amsmath,amssymb]{revtex4-1}
\usepackage{graphicx,dcolumn,bm,times}

\begin{document}


\title{Compressed Exponential Relaxation as Superposition of Dual Structure\\ in Pattern Dynamics of Nematic Liquid Crystals}

\author{T. Narumi}
\affiliation{Department of Applied Quantum Physics and Nuclear Engineering, Kyushu Univ., Fukuoka 819-0395, Japan}
\author{F. Nugroho}
\affiliation{Physics Department, Gadjah Mada Univ., Yogyakarta 55281, Indonesia}
\author{J. Yoshitani}
\affiliation{Department of Applied Quantum Physics and Nuclear Engineering, Kyushu Univ., Fukuoka 819-0395, Japan}
\author{Y. Hidaka}
\affiliation{Department of Applied Quantum Physics and Nuclear Engineering, Kyushu Univ., Fukuoka 819-0395, Japan}
\author{M. Suzuki}
\affiliation{Department of Applied Quantum Physics and Nuclear Engineering, Kyushu Univ., Fukuoka 819-0395, Japan}
\author{S. Kai}
\affiliation{Department of Applied Quantum Physics and Nuclear Engineering, Kyushu Univ., Fukuoka 819-0395, Japan}

\begin{abstract}
Soft-mode turbulence (SMT) is the spatiotemporal chaos observed in homeotropically aligned nematic liquid crystals, 
where non-thermal fluctuations are induced by nonlinear coupling between the Nambu--Goldstone and convective modes.
The net and modal relaxations of the disorder pattern dynamics in SMT have been studied 
to construct the statistical physics of nonlinear nonequilibrium systems.
The net relaxation dynamics is well-described  by a compressed exponential function
and the modal one satisfies a dual structure, dynamic crossover accompanied by a breaking of time-reversal invariance.
Because the net relaxation is described by a weighted mean of the modal ones with respect to the wave number,
the compressed-exponential behavior emerges as a superposition of the dual structure.
Here, we present experimental results of the power spectra  
to discuss the compressed-exponential behavior and the dual structure 
from a viewpoint of the harmonic analysis.
We also derive a relationship of the power spectra from the evolution equation of the modal autocorrelation function.
The formula will be helpful to study non-thermal fluctuations in experiments such as the scattering methods.
\end{abstract}


\maketitle


\section{Introduction}
Nonlinear phenomena are ubiquitous in nature and have various aspects. 
An example is {\it chaotic} (synonymously, {\it turbulent}) dynamics, 
in which few degrees of freedom (d.f.) generate unpredictable behavior (e.g., Refs.~\cite{Strogatz1994, Mori1998_book}).
In spatially extended systems where the number of d.f. increases with expanding the system size, 
weak nonlinearity can lead to spatial and temporal disorder. 
The phenomenon triggered by weak nonlinearity is called spatiotemporal chaos. 
There are several theoretical descriptions (Ref.~\cite{Cross1993} and references therein) 
such as the Kuramoto--Sivashinsky (KS) equation \cite{Kuramoto1984}, 
the coupled map lattices \cite{Kaneko1985},
the Swift--Hohenbarg equation \cite{Chate1987},
the Nikolaevskii equation \cite{Nikolaevskii1989},
and the complex Ginzburg--Landau equation \cite{Chate1994}.
Nevertheless, systematical understanding for the chaotic or turbulent disorder is still developing.

An experimentally obtained spatiotemporal chaos in homeotropically aligned nematic liquid crystals
is soft-mode turbulence (SMT) \cite{Kai1996, Hidaka1997_PRE, Hidaka1997_JPSJ},
induced by nonlinear coupling between the local convective mode and global Nambu--Goldstone one (see Appendix for details).
SMT has a characteristic spatial structure.
Because a patchwork-like structure appears on the $x$-$y$ plane 
when the angle of the local convective mode is color-coded, 
the structure is called patch structure \cite{Tamura2002, Hidaka2006, Tamura2006}.
The characteristic size $\xi$ of the single patch domain is several of convective rolls
and decreases as the distance from the onset of SMT as $\xi\sim\varepsilon^{-1/2}$,
where $\varepsilon = \left(V/V_{c}\right)^{2}-1$ is the normalized voltage used as a control parameter for SMT,
$V$ denotes the magnitude of the applied AC voltage, and $V_{\text{c}}$ the threshold voltage of the electroconvection.

We have investigated SMT dynamics by measuring {\it net} temporal autocorrelation function $\hat{Q}(\tau)$ of turbulence-like dynamics in a steady state.
It had long been considered that the simple exponential described the relaxation dynamics.
However, we revealed that the relaxation deviates from the simple exponential at the vicinity of the SMT's onset \cite{Nugroho_SMTtotal_2012};
instead, it is well-fitted by the so-called Kohlrausch--Williams--Watts (KWW) function \cite{Kohlrausch1854, Williams1970}
\begin{equation}
	\hat{Q}(\tau) = \alpha \exp\left[-\left(\frac{\tau}{\tau_{0}}\right)^{\beta}\right],
\end{equation}
employed to explain the relaxation dynamics in glass-forming liquids (GFLs).
It is considered that the cooperative rearranging in GFLs leads to the KWW-type relaxation \cite{Bouchaud2008};
indeed, spatially and temporally fluctuating domains have been experimentally observed near the glass transition point.
The phenomenon is called the dynamical heterogeneity \cite{Sillescu1999, Ediger2000, Richert2002, Berthier2011_book}. 
The characteristic length of the dynamical heterogeneity increases
as the system approaches the glass transition point.
Although there are many studies for GFLs showing $\beta < 1$ (i.e., stretched exponential),
the compressed-exponential behavior (i.e., $\beta>1$) has been observed in some GFLs (e.g., Refs.~\cite{Cipelletti2000, Chung2006, Fluerasu2007, Caronna2008, Guo2011}).
In particular, Caronna et al. reported that the KWW exponent $\beta$ increases toward $2$ 
with approaching the glass transition point \cite{Caronna2008}.
In SMT, the exponent is unity (i.e., simple exponential) at a large $\varepsilon$
and increases toward $2$ (i.e., Gaussian) with decreasing $\varepsilon$.
The non-exponential behavior of the SMT's net relaxation originates from the patch structure.
We have thus proposed a similarity between dynamics of SMT and GFL from a viewpoint of the spatial structure \cite{Nugroho_SMTtotal_2012}. 

The temporal correlations of each wave number, i.e., \textit{modal} autocorrelation functions $\hat{U}_{k}(\tau)$, 
are also suitable for studying chaotic or turbulent dynamics.
Mori and Okamura have numerically and analytically discovered a dual structure in the KS equation \cite{Mori2009};
the relaxation dynamics is dominated by the deterministic and stochastic orbits 
in the short-time and long-time region, respectively. 
We experimentally found the dual structure of the modal autocorrelation in SMT 
and have theoretically revealed the mechanisms of the dual structure \cite{Narumi_SMTmem_tobe}.
It has been reported that the spatiotemporal disorder in SMT generates non-thermal fluctuations \cite{Tamura2002, Hidaka2010}
by which a non-Markovian (i.e., memory) effect is expected to play a significant role in the relaxation dynamics.
We thus derived an evolution equation for the modal autocorrelation function 
in the projection-operator method for chaos and turbulence \cite{Mori2001}
and have specified the memory effect due to non-thermal fluctuations.
The numerically obtained memory function from the experimental results has suggested that 
the non-thermal fluctuations are separated into Markov and non-Markov parts,
where the latter part was named turbulent fluctuations.
The relaxation dynamics is consequently separated into three relaxation stages: bare-friction stage, early stage, and late stage.
The bare-friction stage disappears
when the dissipation by the turbulent transport has much influence compared to that by the molecular one;
subsequently, the modal relaxation in SMT is represented as
\begin{equation}
	\hat{U}_{k}(\tau) \propto \left\{ \begin{array}{ll}
		1- \left(\tau/ \tau^{(\text{a})}_{k} \right)^{2} & (\tau\ll\tau^{(\Gamma)}_{k}\text{: early stage}) \\
		 & \\
		\exp\left[-\tau\left/ \tau^{(\text{e})}_{k}\right.\right] & (\tau^{(\Gamma)}_{k}\ll\tau\text{: late stage}).
	\end{array} \right. 
\end{equation}
where $\tau^{(\Gamma)}_{k}$ denotes the characteristic timescale of the memory due to the turbulent fluctuations.
Furthermore, we have proved that the memory effect due to the turbulent fluctuations originates from the patch structure \cite{Narumi_SMTmem_tobe}.

In this proceedings paper, 
we uniformly argue the net and modal relaxation dynamics of SMT
and discuss the compressed exponential behavior and the dual structure
from a viewpoint of the harmonic analysis.
This is useful for experiments such as the scattering methods.
The contents are organized as follows.
In Sec.~\ref{sec:statistical}, the net and modal autocorrelation functions are defined
and the memory function is reviewed according to Ref.~\cite{Narumi_SMTmem_tobe}.
In Sec.~\ref{sec:RandD}, we show our experimental results of the power spectra and present the analytical relationship.
We conclude this proceedings paper in Sec.~\ref{sec:summary}.
In Appendix, the detail of SMT and the numerical algorithm for solving the memory function are summarized.

\section{Relaxation Dynamics} \label{sec:statistical}
\subsection{Net and modal correlation functions}
We observed the pattern dynamics of SMT through the transmitted light intensity $I(\vec{r},t)$ with $\vec{r}=(x,y)$.
Let $\hat{Q}(\vec{r},\tau)$ denote the autocorrelation function of the transmitted light intensity,
defined as
\begin{equation}
	\hat{Q}(\vec{r}, \tau) = \frac{\left<\Delta I(\vec{r},t+\tau)\Delta I(\vec{r},t)\right>}{\left<\Delta I(\vec{r},t)^{2}\right>},
	\label{fluc_cor}
\end{equation}
where $\Delta I(\vec{r},t) = I(\vec{r},t) - \left<I(\vec{r},t)\right>$ denotes the fluctuation of the transmitted light intensity
and the angle brackets indicate the long-time average in steady state;
\begin{equation}
	\left<f(t+\tau)g(t)\right> = \lim_{T\to \infty}\frac{1}{2T}\int_{-T}^{T}{\rm d}t f(t+\tau)g(t).
	\label{def_LTaverage}
\end{equation}
The net autocorrelation function $\hat{Q}(\tau)$ can be defined as the spatial average of $\hat{Q}(\vec{r},\tau)$;
\begin{equation}
	\hat{Q}(\tau) = \frac{1}{V_{d}} \int_{V_{d}}{\rm d}\vec{r}~\hat{Q}(\vec{r},\tau),
	\label{total_cor}
\end{equation}
where $V_{d}$ denotes the generalized volume of $d$-dimensional system and the integral range is whole system.
The net relaxation in SMT is well described by the KWW function
and the KWW exponent $\beta$ is $1$ at a large $\varepsilon$ and approaches to $2$ with decreasing $\varepsilon$ \cite{Nugroho_SMTtotal_2012}.

In addition, we have focused on the modal element $u_{\vec{k}}(t)$ of the fluctuation $\Delta I(\vec{r},t)$;
\begin{equation} 
	u_{\vec{k}}(t) = \int_{V_{d}} {\rm d} \vec{r} \Delta I(\vec{r},t)e^{\text{i} \vec{k} \cdot \vec{r}}, 
\end{equation}
where $\text{i}=\sqrt{-1}$.
Because SMT is isotropic \cite{Kai1996}, it is sufficient to study $u_{k}(t)$,
where $k$ denotes the radial wave number; $k=|\vec{k}|$.
The modal autocorrelation function is defined as
\begin{equation}
	\hat{U}_{k}(\tau) = \frac{\left<u_{k}(t+\tau)u_{k}^{*}(t)\right>}{P_{k}}.
	\label{def_corr_func}
\end{equation}
Here, the asterisk symbol denotes taking the complex conjugate 
and $P_{k} = \left<|u_{k}(t)|^{2}\right>$ the spatial power spectrum.
The modal correlation function $\hat{U}_{k}(\tau)$ is a real number due to the translational symmetry and isotropy.
In the rest of this paper, the wave number $k$ is normalized by $\lambda_{0}$ as $\hat{k}=k \lambda_{0}/2\pi$,
where $\lambda_{0}=\lambda_{0}(\varepsilon)$ denotes the length of a pair of electroconvections.

One can derive the analytical relationship between the net and modal autocorrelation functions.
The autocorrelation of $u_{\vec{k}}(t)$ reduces to
\begin{equation}
\begin{array}{cl}
	 	&	\left<u_{\vec{k}}(t+\tau)u^{*}_{\vec{k}}(t)\right>  \\
		&	\\
	= 	&	\iint _{V_{d}^{2}}{\rm d}\vec{r}{\rm d}\vec{r^{\prime}}
			\left<\Delta I(\vec{r},t+\tau)\Delta I(\vec{r^{\prime}},t)\right>e^{i(\vec{r}-\vec{r^{\prime}})\cdot\vec{k}}.
\end{array}
\end{equation}
Integrating the both sides with respect to $\vec{k}$, we obtain
\begin{equation}
	\hat{Q}(\tau) \propto \int\frac{{\rm d}\vec{k}}{(2\pi)^{d}}P_{k}\hat{U}_{k}(\tau).
\end{equation}
For $d=2$, the net autocorrelation function connects to the modal one as
\begin{equation}
	\hat{Q}(\tau) = \frac{{\displaystyle \int_{0}^{\infty} {\rm d}k~kP_{k} \hat{U}_{k}(\tau)}}{{\displaystyle \int_{0}^{\infty} {\rm d}k~kP_{k}}}.
	\label{total_and_modal}
\end{equation}
Therefore, in SMT, the compressed-exponential behavior appeared in the net relaxation is interpreted 
as a superposition of the dual structure in the modal ones.

\subsection{Memory function} \label{subsec:Mem}
We have proposed an approach by the memory function derived in the projection-operator formalism \cite{Narumi_SMTmem_tobe}.
The evolution equation for the modal autocorrelation function is represented as
\begin{equation}
	\frac{\partial \hat{U}_{k}(\tau)}{\partial \tau} = -\int_{0}^{\tau}{\rm d}\tau^{\prime}\Gamma_{k}^{\prime}(\tau-\tau^{\prime})\hat{U}_{k}(\tau^{\prime}),
	\label{EOM_modalcorr_prime}
\end{equation}
where $\Gamma^{\prime}_{k}(\tau)$ denotes the memory function that results from the non-thermal fluctuations.
The numerically-obtained memory function $\Gamma^{\prime}_{k}(\tau)$ has a sharp peak at $\tau=0$,
implying that the non-thermal fluctuations in SMT can be divided into Markov and non-Markov parts \cite{Narumi_SMTmem_tobe},
i.e., 
\begin{equation}
	\Gamma_{k}^{\prime}(\tau) = 2\gamma^{(0)}_{k}\delta(\tau)+\Gamma_{k}(\tau).
	\label{memory_function_SMT}
\end{equation}
where $\gamma^{(0)}_{k}$ denotes the bare friction due to the Markov part of the non-thermal fluctuations
and $\Gamma_{k}(\tau)$ the memory function caused from the non-Markov part.
The evolution equation \eqref{EOM_modalcorr_prime} reduces to
\begin{equation}
	\frac{\partial \hat{U}_{k}(\tau)}{\partial \tau} =  -\gamma^{(0)}_{k}\hat{U}_{k}(\tau)-\int_{0}^{\tau}{\rm d}\tau^{\prime}\Gamma_{k}(\tau-\tau^{\prime})\hat{U}_{k}(\tau^{\prime}).
	\label{EOM_modalcorr}
\end{equation}
The non-Markov part of the non-thermal fluctuations is called the turbulent fluctuations.
The algorithm to numerically solve the memory function is summarized in Appendix.

The \textit{macroscopic} friction $\gamma^{(\Gamma)}_{k}$ due to the turbulent fluctuations is represented as \cite{Mori2001}
\begin{equation}
	\gamma^{(\Gamma)}_{k} = \int_{0}^{\infty}{\rm d}\tau \Gamma_{k}(\tau).
	\label{gamma_Gamma}
\end{equation}
Moreover, one can define a characteristic time scale of the memory function as
\begin{equation}
	\tau^{(\Gamma)}_{k}  = \int_{0}^{\infty}{\rm d}\tau \frac{\Gamma_{k}(\tau)}{\Gamma_{k}(0)} = \frac{\gamma^{(\Gamma)}_{k}}{\Gamma_{k}(0)},
	\label{tau_Gamma}
\end{equation}
within which the memory effect due to the turbulent fluctuations is alive;
in other words, for $\tau^{(\Gamma)}_{k} \ll \tau$ (late stage), 
the memory function due to the turbulent fluctuations is approximately represented by
\begin{equation}
	\Gamma_{k}(\tau) \simeq 2\gamma^{(\Gamma)}_{k}\delta(\tau).
\end{equation}
In contrast, the non-Markovian regime $\tau \ll \tau^{(\Gamma)}_{k}$ can be analytically divided 
into $\tau \ll \tilde{\gamma}_{k}\tau^{(\Gamma)}_{k}$ (bare-friction stage) 
and $ \tilde{\gamma}_{k}\tau^{(\Gamma)}_{k} \ll \tau \ll \tau^{(\Gamma)}_{k}$ (early stage) \cite{Narumi_SMTmem_tobe},
where $\tilde{\gamma}_{k}$ denotes the ratio of the bare friction to the turbulent friction,
\begin{equation}
	\tilde{\gamma}_{k} = \gamma^{(0)}_{k} \left/ \gamma^{(\Gamma)}_{k}\right. .
\end{equation}
In the case that $\tilde{\gamma}_{k} \ll 1$, the bare-friction stage shrinks;
hence, the early and late stages are observed as the dual structure in SMT.

\section{Harmonic Analysis} \label{sec:RandD}
\begin{figure}
\includegraphics[width=0.99\linewidth]{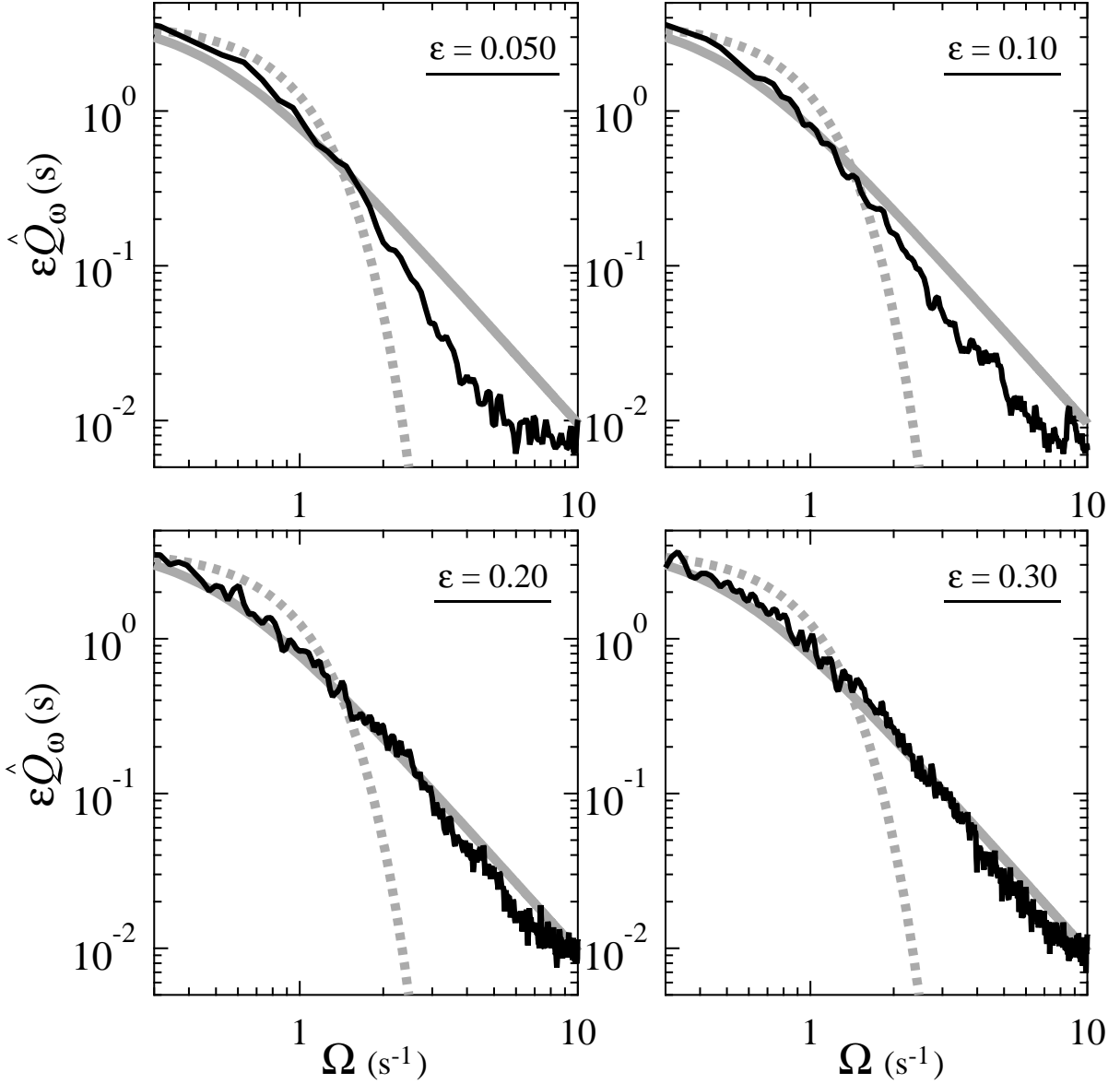}
\caption{
	Log-log plots of the power spectra $\varepsilon Q_{\omega}$ (black line) for several control parameters;
	$\varepsilon = 0.050$ (top left), $0.10$ (top right), $0.20$ (bottom left), and $0.30$ (bottom right). 
	A gray solid line indicates the Lorenzian~\eqref{PS_net_smp_exp} 
	and a gray dotted line the Gaussian~\eqref{PS_net_smp_gauss},
	where $\Lambda$ is set as $0.48$ s$^{-1}$.
}
\label{fig:PS}
\end{figure}
We first present the power spectrum $\hat{Q}_{\omega}$ of the fluctuation of $I(\vec{r},t)$ in Fig.~\ref{fig:PS},
where $\hat{Q}_{\omega}$ was calculated from $\hat{Q}(\tau)$ as
\begin{equation}
	\hat{Q}_{\omega} = 2\int_{0}^{\infty} {\rm d}\tau \hat{Q}(\tau)\cos\omega \tau.
	\label{PS_net}
\end{equation}
In the case that the relaxation satisfies the simple exponential function (i.e., $\beta=1$ in the KWW function),
the power spectrum is analytically derived as
\begin{equation}
	\hat{Q}_{\omega} = \frac{2 \tau_{0}} {1+\left(\tau_{0}\omega\right)^{2}},
	\label{PS_net_smp_exp_origin}
\end{equation}
where $\tau_{0}=\tau_{0}(\varepsilon)$ denotes the characteristic correlation time of the relaxation.
According to Ref.~\cite{Hidaka1997_JPSJ},
the above relationship reduces to 
\begin{equation}
	\varepsilon \hat{Q}_{\omega} = \frac{2\Lambda} {\Lambda^{2}+\Omega^{2}}
	\label{PS_net_smp_exp}
\end{equation}
with $\Lambda^{-1}=\tau_{0}\varepsilon$ and $\Omega = \omega / \varepsilon$.
As discussed in Ref.~\cite{Hidaka1997_JPSJ}, 
the right-hand side of Eq.~\eqref{PS_net_smp_exp} does not depend on the control parameter $\varepsilon$ 
because $\tau_{0}$ is inversely proportional to $\varepsilon$ \cite{Kai1996}.
On the other hand, when the relaxation is the Gaussian (i.e., $\beta=2$ in the KWW function),
the power spectrum is analytically represented as
\begin{equation}
	\varepsilon \hat{Q}_{\omega} = \frac{\sqrt{\pi}}{\Lambda} \exp \left[ - \frac{\Omega^{2}}{4\Lambda ^{2}}\right].
	\label{PS_net_smp_gauss}
\end{equation}
%
To the best of our knowledge, no analytical functions for the Fourier transform of the KWW function with $1 < \beta < 2$ have been proposed yet.
As shown in the top panels of Fig.~\ref{fig:PS}, 
$\varepsilon \hat{Q}_{\omega}$ for a small $\varepsilon$, where $\beta > 1$, 
is in domains bounded by Eqs.~\eqref{PS_net_smp_exp} and \eqref{PS_net_smp_gauss}.
Morishita has proposed that the CEF-type decay appears as coexistence of ballistic and diffusive motion \cite{Morishita2012}.
This plot style, $\varepsilon \hat{Q}_{\omega}$ versus $\Omega$, helps us to intuitively understand the KWW exponent $1 < \beta < 2$.
Further, as shown in the bottom panels of Fig.~\ref{fig:PS},
Eq.~\eqref{PS_net_smp_exp} well describes the power spectrum at a higher $\varepsilon$.
It is evident that $\beta$ converges to unity with increasing $\varepsilon$.

\begin{figure}[!t]
\begin{center}
\includegraphics[width=0.99\linewidth]{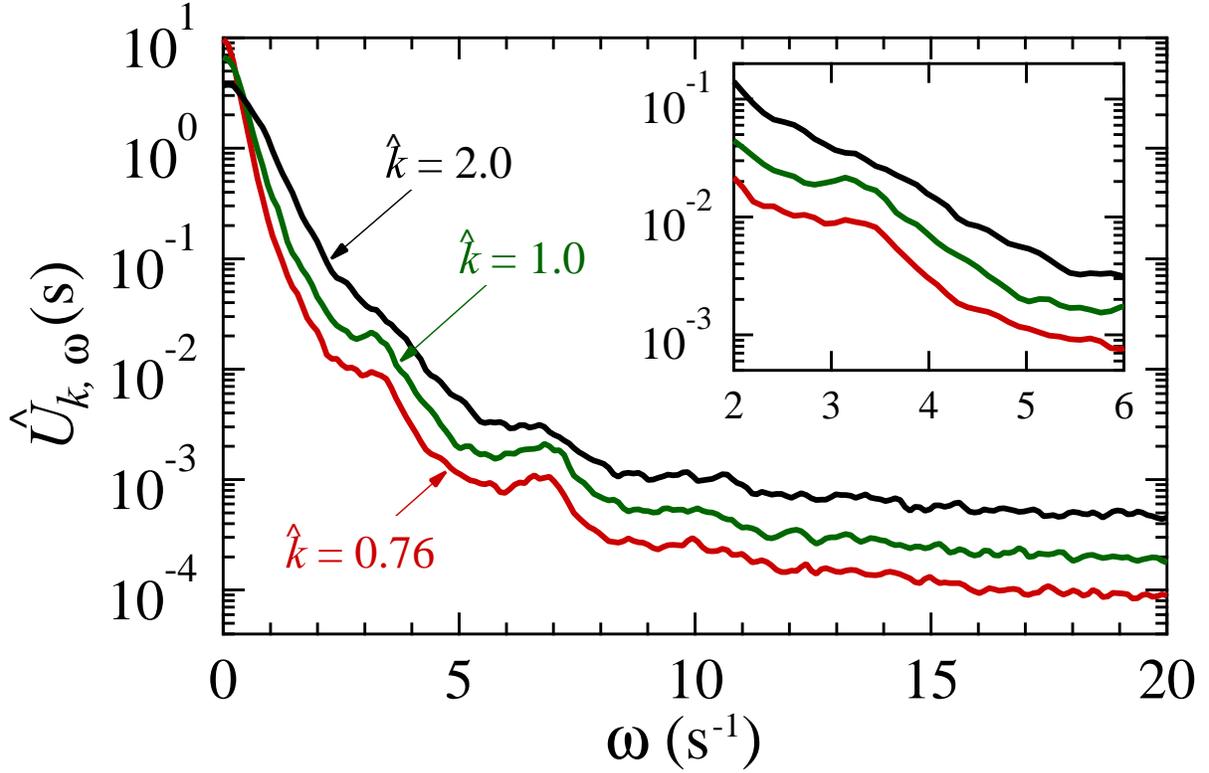}
\end{center}
	\caption{
		Plot of the power spectrum $\hat{U}_{k,\omega}$ at $\varepsilon=0.10$
		for several wave numbers;
		$\hat{k}=0.76$, $1.0$, and $2.0$ from bottom to top.
		The inner plot is a closeup in an intermediate regime.
	}
\label{fig:FT_modal}
\end{figure}
Next, we show results of the power spectrum $\hat{U}_{k,\omega}$ of the modal elements in Fig.~\ref{fig:FT_modal},
where $\hat{U}_{k,\omega}$ was calculated by
\begin{equation}
	\hat{U}_{k,\omega} = 2\int_{0}^{\infty} {\rm d}\tau \hat{U}_{k}(\tau) \cos \omega \tau.
\end{equation}
The harmonic analysis emphasizes on the short-time dynamics.
Indeed, the results support appearance of the early and bare-friction stages in SMT dynamics:
As shown in the inner plot of Fig.~\ref{fig:FT_modal},
the spectra decay linearly in the semi-log plot within the intermediate frequency regime,
signifying that the modal correlation function is well-described by the algebraic function in the early stage.
They gently decay at a high frequency,
implying the existence of an extremely short-time dynamics (i.e., the bare-friction stage) 
dominated by the Markov part of the non-thermal fluctuations.
Note that the power spectra of the KS equation exponentially decay in large-frequency regime \cite{Mori2009}
because the bare-friction stage does not exist there. 

\begin{figure}[!t]
\begin{center}
\includegraphics[width=0.99\linewidth]{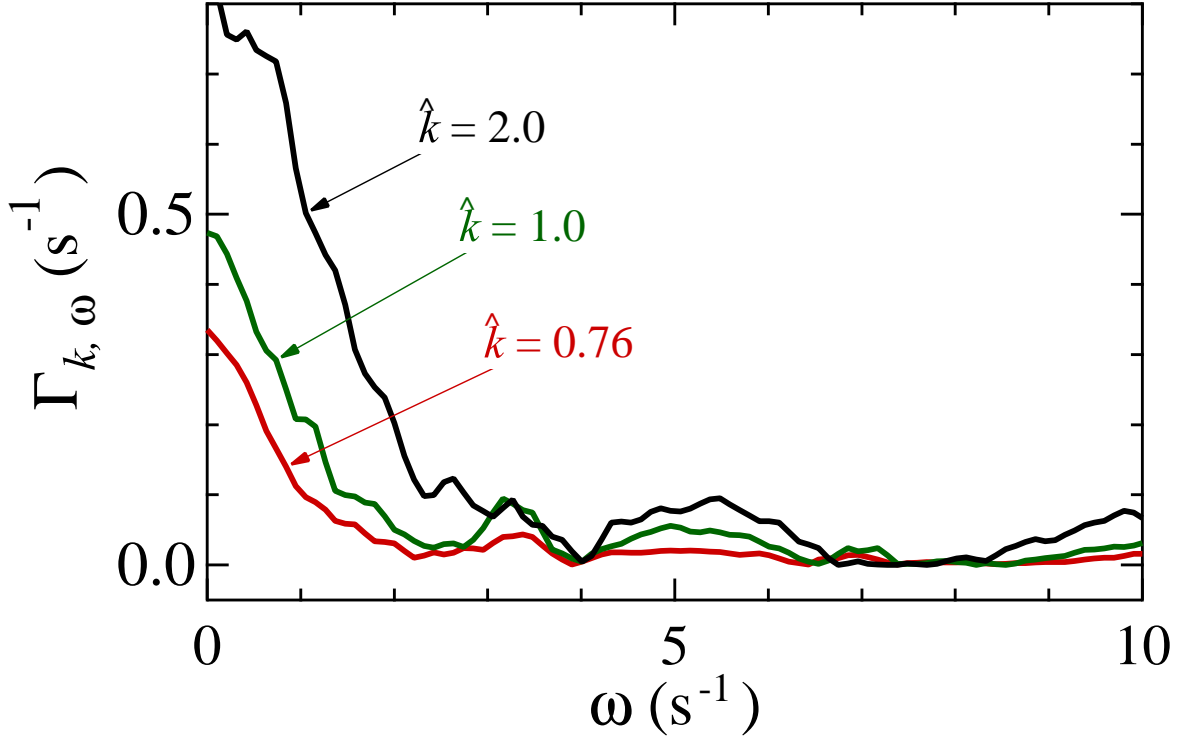}
\end{center}
	\caption{
		Plot of the power spectrum $\Gamma_{k,\omega}$ at $\varepsilon=0.10$
		for several wave numbers;
		$\hat{k}=0.76$, $1.0$, and $2.0$ from bottom to top.
	}
\label{fig:FT_mem}
\end{figure}
Finally, the Fourier transform $\Gamma_{k.\omega}$ of the memory function due to the turbulent fluctuations is shown in Fig.~\ref{fig:FT_mem},
where $\Gamma_{k,\omega}$ was calculated by
\begin{equation}
	\Gamma_{k,\omega} = 2\int_{0}^{\infty} {\rm d}\tau \Gamma_{k}(\tau) \cos \omega \tau.
\end{equation}
The power spectrum $\Gamma_{k,\omega}$ analytically relates to $\hat{U}_{k,\omega}$.
Corresponding to Eq.~\eqref{EOM_modalcorr_prime},
the evolution equation of $u_{k}(t)$ can be represented as
\begin{equation}
	\frac{\partial u_{k}(t)}{\partial t} 	= -\int_{-\infty}^{t}{\rm d} t^{\prime} \Gamma^{\prime}_{k}(t-t^{\prime}) u_{k}(t^{\prime}) + R^{\prime}_{k}(t),
	\label{EOM_mode}
\end{equation}
where $R^{\prime}_{k}(t)$ denotes the non-thermal fluctuations and satisfies 
\begin{equation}
	\left<R^{\prime}_{k}(t)u^{*}_{k}(t_{0})\right> = 0.
\end{equation}
Note that the integral range of Eq.~\eqref{EOM_mode} has been expanded compared to Eq.~\eqref{EOM_modalcorr_prime};
nevertheless, the contribution is included in the fluctuating term \cite{Kubo1991_book}.
In the theoretical framework proposed in Ref.~\cite{Mori2001},
the following relation holds;
\begin{equation}
	\Gamma^{\prime}_{k}(\tau)  = \frac{\left<R^{\prime}_{k}(t+\tau) R^{\prime}_{k}(t)\right> }{P_{k}},
	\label{FDT_nonthermal} 
\end{equation}
indicating that $\Gamma^{\prime}_{k,\omega}$ allows us to investigate the non-thermal fluctuations.
It should be noted here that Eq.~\eqref{FDT_nonthermal} holds in non-equilibrium systems.
From Eq.~\eqref{EOM_mode}, the Fourier transform $u_{k,\omega}$ of $u_{k}(t)$ satisfies
\begin{equation}
	\left(\text{i} \omega + \Gamma^{\prime}_{k}[\omega] \right) u_{k,\omega}  = R^{\prime}_{k,\omega},
	\label{FT_EOM_mode}
\end{equation}
where $\Gamma^{\prime}_{k}[\omega]$ denotes the Fourier-Laplace transform of $\Gamma^{\prime}_{k}(t)$, 
defined as
\begin{equation}
	\Gamma^{\prime}_{k}[\omega]  = \int_{0}^{\infty}{\rm d}\tau \Gamma^{\prime}_{k}(\tau) e^{-\text{i}\omega \tau}.
	\label{FLT_mem}
\end{equation}
The power spectrum of $u_{k}(t)$ is thus written as
\begin{equation}
	\hat{U}_{k,\omega} = \frac{{\displaystyle \Gamma^{\prime}_{k,\omega}}}{{\displaystyle |\text{i}\omega+\Gamma^{\prime}_{k}[\omega]|^{2}}}.
	\label{PS_mode}
\end{equation}
Under the condition \eqref{memory_function_SMT}, Eq.~\eqref{PS_mode} reduces to
\begin{equation}
	\hat{U}_{k,\omega} = \frac{{\displaystyle 2\gamma^{(0)}_{k} + \Gamma_{k,\omega}}}{{\displaystyle \left(\gamma^{(0)}_{k}+\left. \Gamma_{k,\omega}\right/ 2\right)^{2} + \left(\omega+\text{Im}\Gamma_{k}[\omega]\right)^{2}}}.
	\label{PS_mode_mem}
\end{equation}
Although this is equivalent to Eq.~\eqref{EOM_modalcorr},
it might be more compatible with experiments such as the scattering methods.
Because $\Gamma_{k,\omega}$ rapidly decay as shown in Fig.~\ref{fig:FT_mem},
Eq.~\eqref{PS_mode_mem} for a sufficiently large $\omega$ approximately reduces to
\begin{eqnarray}
	\frac{\partial^{2} \hat{U}_{k}(\tau)}{\partial \tau^{2}} & = & - 2\gamma^{(0)}_{k}\delta(\tau) \cr
	& & \cr
	\Rightarrow~~\hat{U}_{k}(\tau) & = & 1-\gamma^{(0)}_{k} |\tau|,
\end{eqnarray}
i.e., the relaxation in the bare-friction stage is approximately linear decay, 
as mentioned in Ref.~\cite{Narumi_SMTmem_tobe}.
In contrast, at a sufficiently small $\omega$ where $\text{Im}\Gamma_{k}[\omega]\to 0$,
we obtain the macroscopic friction coefficient \cite{Mori2001} as
\begin{equation}
	\gamma^{(0)}_{k} +\gamma^{(\Gamma)}_{k} = 1/\tau^{(\text{U})}_{k},
\end{equation}
where $\tau^{(\text{U})}_{k}$ denotes the characteristic timescale of the modal autocorrelation function,
defined as
\begin{equation}
	\tau^{(\text{U})}_{k} = \int_{0}^{\infty}{\rm d}\tau \hat{U}_{k}(\tau).
\end{equation}
%

\section{Summary} \label{sec:summary}
We have studied the relaxation dynamics of SMT, 
aiming to clarify statistical-physical properties in the nonlinear nonequilibrium systems.
Because the \textit{net} autocorrelation function is a weighted mean of the \textit{modal} autocorrelation functions,
it has turned out that the CEF decay in SMT appears as a superposition of the dual structure.
Further, this proceedings paper has concentrated on the temporal power spectra of pattern dynamics and its modal element.
The plot style of Fig.~\ref{fig:PS} is compatible with the viewpoint 
that the coexistence of the ballistic and diffusive motions produces a CEF decay.
The physical origin of the relationship between CEF and complex systems such as the glassy dynamics is still an open question;
however, the emergence of the CEF decay on SMT will provide us with a clue.
In addition, we have shown the power spectra $\hat{U}_{k,\omega}$ and $\Gamma_{k,\omega}$
and discussed the dual structure from a viewpoint of the harmonic analysis.
Techniques such as the scattering experiments can measure $\gamma^{(0)}+\Gamma_{k}[\omega]$ as the complex friction coefficients;
therefore, the equations such as Eq.~\eqref{PS_mode_mem} will be helpful to investigate the turbulent fluctuations.

\begin{acknowledgments}
The authors gratefully acknowledge Prof.~Tomoyuki Nagaya (Oita University) for productive suggestions.
This work was partially supported by KAKEHNI (Nos.~21340110 and 24540408),
a Grant-in-Aid for Scientific Research on Innovative Areas---"Emergence in Chemistry"~(No.~20111003),
and the JSPS Core-to-Core Program "International research network for non-equilibrium dynamics of soft matter."
\end{acknowledgments}

\appendix

\section*{Appendix}

\subsection*{Soft-mode turbulence} \label{sec:experiment}
Instability in liquid crystal systems leads to pattern formation \cite{Chandrasekhar1992, deGennes1995}. 
Electrohydrodynamic convection is such a self-organizing phenomenon observed in nematic liquid crystal \cite{Williams1963}.
Let us consider systems of nematic liquid crystals with negative dielectric constant anisotropy $\epsilon_{\text{a}}= \epsilon_{\parallel} - \epsilon_{\perp}$, 
where $\epsilon_{\parallel}$ denotes the dielectric constant parallel to the director $\textbf{n} = \textbf{n}(x,y,z)$
representing the molecular orientation of a liquid crystal,
$\epsilon_{\perp}$ the dielectric constant perpendicular to $\textbf{n}$.
When the AC voltage larger than a threshold magnitude is applied to the systems,
the electroconvection occurs due to the so-called Carr--Helfrich effect \cite{Carr1969, Helfrich1969},
interaction between the anisotropy of the liquid crystals and electrical current by impurity ions.
The characteristic length scale of the electroconvection is much shorter than that of the heat convection,
and consequently characteristic time scales of the electroconvection are easy to access in experiments.
Thus, the electrohydrodynamic convection has been studied as an ideal subject for the dissipative structure \cite{Kai1991_Proceeding}.
The squared magnitude $V^{2}$ and inverse frequency $1/f$ of the AC voltage correspond 
to the Rayleigh number and the Prandtl number, respectively \cite{Kai1976},
where the former is proportional to the temperature difference 
and the latter is equivalent to the ratio of the viscous diffusivity to the thermal one
in the Rayleigh--B\'{e}nard convection.

There are two types of layer alignment in nematic liquid crystal;
one is planar alignment in which the director aligns parallel to substrates ($x$-direction),
and the other is homeotropic alignment in which the director aligns perpendicular to substrates ($z$-direction).
The rubbing to $x$-direction on the substrate's surface ($x$--$y$ plane) is treated to make planer systems,
and intrinsically breaks the rotational symmetry.
In systems where nematic liquid crystals are homeotropically anchored at boundary ($x$-$y$ plane),
rotational symmetry are alive at the ground state as well as translation symmetry.
With increasing magnitude $V$ of the AC voltage parallel to $z$-axis,
the Fr\'{e}edericksz transition occurs at a threshold voltage $V_{\text{F}}$ \cite{Freedericksz1933}.
The directors $\textbf{n}$ tilt in an arbitrary direction with respect to the $z$-axis above $V_{\text{F}}$,
accompanied by spontaneous violation of the rotational symmetry.
The projection of the director onto the $x$-$y$ plane, called $\textbf{C}$-director, 
can rotate on the $x$-$y$ plane without requiring additional energy, 
i.e., $\textbf{C}$-director behaves as the Nambu--Goldstone mode \cite{Hertrich1992, Richter1995_EL, Tribelsky1996}.
With further increasing $V$, 
the electrohydrodynamic convection occurs at the electroconvective threshold voltage $V_{\text{c}}$.
The nonlinear coupling between the convective mode $\textbf{q}(x,y)$ and the Nambu--Goldstone one $\textbf{C}(x,y)$ induces 
spatially and temporally disordered pattern called soft-mode turbulence (SMT) \cite{Kai1996, Hidaka1997_PRE, Hidaka1997_JPSJ}.
SMT is regarded as an experimentally observed spatiotemporal chaos.
The diversity of the relaxation time at the onset of SMT implies that SMT arises supercritically at $V_{\text{c}}$.
The correlation length is finite but much longer than excited and dissipative lengths,
where energy is injected at the excited length scale and dissipated at the smaller one.

There are two types of SMT pattern;
oblique rolls in $f < f_{\text{L}}$ and normal rolls in $f > f_{\text{L}}$,
where $f_{\text{L}}$ denotes the Lifshitz frequency \cite{Kai1996,Hidaka1997_PRE}.
The interaction between the convective and Goldstone modes are different between oblique and normal rolls \cite{Rossberg1996, Rossberg1997};
the convective wave vector in SMT tend to become parallel to the $\textbf{C}$-director in the normal roll and oblique in the oblique roll.
The patch structure is observed only in the oblique rolls \cite{Hidaka2006, Tamura2006}.

We have studied the two-dimensional pattern dynamics of SMT.
The experimental setup for this study refers to a standard one \cite{Kai1996, Tamura2001, Anugraha2008, Nugroho_SMTtotal_2012, Narumi_SMTmem_tobe}.
We filled the space between two parallel glass plates with the nematic liquid crystal, $N$--(4--Methoxybenzilidene)--4--buthylaniline (MBBA).
The surfaces of the plates were coated with transparent electrodes, indium tin oxide (ITO),
which is a circular electrode with the radius $13$ mm.
The dielectric constant parallel to the director was $6.25$
and the electric conductivity parallel to the director $1.17 \times 10^{-7}~\Omega^{-1}\text{m}^{-1}$.
Note that the dielectric constant anisotropy $\epsilon_{\text{a}}$ is negative.
The thickness between the plates is $50~\mu$m for the net relaxation dynamics and $27~\mu$m for the modal relaxation dynamics.
To obtain the homeotropic alignment,
the surface was laid by a surfactant,  DMOAP ($N$, $N$--dimethyl--$N$--octadecyl--3--aminopropyl--trimethoxysilyl chloride 50\%). 

AC voltage $V(t)=\sqrt{2}V\cos(2\pi f t)$ was applied to the sample.
The threshold voltage $V_{\text{c}}$ for electroconvection was $7.78 \pm 0.05$~V. 
We fixed $f=100$~Hz that was much less than $f_{\text{L}}$ to obtain the oblique rolls.
The temperature was set at $30.00\pm 0.05~^{\circ}\text{C}$ and the magnetic field was absent in this study.
Before each sampling,
we waited for $10$~min at $V_{\text{w}}=6.0$~V and then for $10$~min at set $V$, 
where $V_{\text{F}}<V_{\text{w}}<V_{\text{c}}$.
The waiting time is sufficiently long for systems to become steady state.

The transmitted light intensity at each pixel were digitized into $8$-bit (i.e., $256$-level) information,
where a series of pattern analyzing was processed according to Ref.~\cite{Nagaya1999}.

\subsection*{Numerical algorithm to calculate memory function due to turbulent fluctuations}
In order to solve the memory function numerically 
by using an experimentally-obtained modal autocorrelation function $\hat{U}(\tau)$, 
we employed fitting by an expansion for short-time regime 
and numerical integral of the evolution equation \eqref{EOM_modalcorr} for otherwise.
Note that the wave-number dependence was omitted in the appendix;
however, the coefficients actually depend on the wave number $k$.

We first obtained the coefficients $\{a_{n}\}$ of an expansion around $\tau=0$ by fitting:
\begin{equation}
	\hat{U}(\tau) = 1 + \sum_{n=1}^{\infty} a_{n}\tau^{n}.
	\label{Taylor_cor}
\end{equation}
Suppose that the memory function $\Gamma(\tau)$ caused by the turbulent fluctuations is described by a continuous function,
i.e.,
\begin{equation}
	\Gamma(\tau) =\sum_{n=0}^{\infty} b_{n}\tau^{n}.
	\label{Taylor_mem}
\end{equation}
Equation \eqref{EOM_modalcorr} provides the bare friction as
\begin{equation}
		\gamma^{(0)} = -a_{1}.
\end{equation}
Moreover, the coefficients $\{b_{n}\}$ of the memory function as
\begin{equation}
	b_{0} = {a_{1}}^{2}-2a_{2}
	\label{Taylor_coef_mem_0}
\end{equation}
and for $n\ge 1$,
\begin{eqnarray}
	b_{n} 	& = & -(n+2)(n+1)a_{n+2} + (n+1)a_{1}a_{n+1} \cr
			&	& \cr
			&	& -\sum_{m=1}^{n}a_{m}b_{n-m} \left(\begin{array}{c} n \\ m \end{array}\right)^{-1}.
	\label{Taylor_coef_mem_n}%
\end{eqnarray}

For the long-time region, we solved the difference equation of Eq.~\eqref{EOM_modalcorr}.
We first introduce $\Delta U_{N}$ as the central difference;
\begin{equation}
	\Delta \hat{U}_{N} = \frac{\hat{U}_{N+1/2} - \hat{U}_{N-1/2}}{\Delta t}
\end{equation}
where $\hat{U}_{N}=\hat{U}(N\Delta t)$, $N$ denotes an integer, and $\Delta t$ the time lag.
Then, the evolution equation is digitalized to
\begin{eqnarray}
	\Delta \hat{U}_{N+1/2} 	& = & \frac{\hat{U}_{N+1} - \hat{U}_{N}}{\Delta t} \cr
							& & \cr
							& = & -\gamma^{(0)}\hat{U}_{N+1/2} - \int_{0}^{\tau}{\rm d}\tau^{\prime}\Gamma(\tau^{\prime})\hat{U}(\tau - \tau^{\prime}), \cr
							& &
\end{eqnarray}
where $\tau$ should be interpreted as $(N+1/2)\Delta t$.
The convolutional integral is further digitalized to
\begin{eqnarray}
	&    & \int_{0}^{\tau}{\rm d}\tau^{\prime}\Gamma(\tau^{\prime})\hat{U}(\tau - \tau^{\prime}) \cr	&    &\cr
	& = & \int_{0}^{\Delta t/2}{\rm d}\tau^{\prime}\Gamma(\tau^{\prime})\hat{U}(\tau - \tau^{\prime}) \cr
	&    & + \sum_{M=1}^{N}\int_{(M-1/2)\Delta t}^{(M+1/2)\Delta t}{\rm d}\tau^{\prime}\Gamma(\tau^{\prime})\hat{U}(\tau - \tau^{\prime}) \cr
	&    &\cr
	& \simeq & \frac{\Delta t}{2}\Gamma_{0}\hat{U}_{N+1/2} + \sum_{M=1}^{N} \Delta t \Gamma_{M}\hat{U}_{N - M + 1/2} 
\end{eqnarray}
with $\Gamma_{N}=\Gamma(N\Delta t)$.
Therefore, the memory function is solved numerically as
\begin{eqnarray}
	\Gamma_{N}	& = & -\frac{2}{\Delta t^{2}}\frac{\hat{U}_{N+1}-\hat{U}_{N}}{\hat{U}_{1}+\hat{U}_{0}} - \frac{\gamma^{(0)}}{\Delta t}\frac{\hat{U}_{N+1}+\hat{U}_{N}}{\hat{U}_{1}+\hat{U}_{0}} \cr
				&    &\cr
				&    & -\frac{\Gamma_{0}}{2}\frac{\hat{U}_{N+1}+\hat{U}_{N}}{\hat{U}_{1}+\hat{U}_{0}} \cr
				&    &\cr
				&    & -\frac{1}{\hat{U}_{1}+\hat{U}_{0}} \sum_{M=1}^{N-1}\Gamma_{M}(\hat{U}_{N - M +1} + \hat{U}_{N - M})
				\label{diff_memory}
\end{eqnarray}
for $N\ge 1$.
The above discretization is not unique, of course;
however, the way to discretize contributes only higher-order errors.

The assumption \eqref{Taylor_mem} does not hold for $\Gamma^{\prime}(\tau)$
because it has singularity at $\tau \simeq 0$ originates from the Markov part of the non-thermal fluctuations.
When one solves $\Gamma^{\prime}(\tau)$ numerically,
the discrete analysis \eqref{diff_memory} with $\gamma^{(0)}=0$ should be used for whole time region.
Note that the initial value of the memory function  is described by
\begin{equation}
	\Gamma^{\prime}_{0} = -\frac{4}{\Delta t^{2}}\frac{\hat{U}_{1}-\hat{U}_{0}}{\hat{U}_{1}+\hat{U}_{0}}.
\end{equation}
%



\end{document}